\documentclass{aa}  
\usepackage{graphicx}
\usepackage{longtable}
\usepackage{tabularx}
\usepackage{txfonts}
\usepackage{hyperref}
\usepackage{amsmath}
\usepackage{scalerel}
\usepackage{natbib}
\usepackage{xcolor}
\definecolor{dark-red}{rgb}{0.9,0.0,0.0}
\definecolor{dark-blue}{rgb}{0.15,0.15,0.9}
\definecolor{dark-green}{rgb}{0.15,0.8,0.15}
\definecolor{medium-blue}{rgb}{0,0,0.9}
\hypersetup{
    colorlinks, linkcolor=red,
    citecolor={dark-blue} , urlcolor={medium-blue}
}
\newcommand{\mjup}{$M_{\rm J}$\,}

\newcommand{\gaia}{{\small\texttt{Gaia}}\,}

\newcommand{\fwfs}{$f_{\scaleto{\texttt{WFS}}{3pt}}$\,}
\newcommand{\wfs}{\texttt{WFS}}
\newcommand{\fwhm}{\texttt{FWHM}}
\newcommand{\psf}{\texttt{PSF}}
\newcommand{\sr}{\texttt{SR}}
\newcommand{\tcatone}{\texttt{TCAT10}\,}
\newcommand{\tcatthree}{\texttt{TCAT30}\,}
\newcommand{\rct}{RC$_\texttt{300}$\,}

\begin{document}

   \title{SPHERE adaptive optics performance for faint targets}

%   \subtitle{}

   \author{M. I. Jones \inst{1}
   \and{J. Milli} \inst{2}
   \and{I. Blanchard} \inst{1}
   \and{Z. Wahhaj} \inst{1}
   \and{R. de Rosa} \inst{1}
   \and{C. Romero}  \inst{1}
%   \and{B. Courtney-Barrer} \inst{1}
%   \and{R. Kokotanekova} \inst{1}
%   \and{J. Kolb} \inst{1}}
}
   \institute{European Southern Observatory, Alonso de C\'ordova 3107, Vitacura, Casilla 19001, Santiago, Chile \\\email{mjones@eso.org}
   \and{Institute of Planetology and Astrophysics of Grenoble (IPAG), Universit\'e Grenoble Alpes}
             }

   \date{}

% \abstract{}{}{}{}{} 
% 5 {} token are mandatory
 
  \abstract
   {High contrast imaging is a powerful technique to detect and characterize planetary companions at large orbital separations from their parent stars.} 
   {We aim at studying the limiting magnitude of the VLT/SPHERE Adaptive Optics  system and the corresponding instrument performance for  faint targets (G $\ge$ 11.0 mag).}
   {We computed coronagraphic H-band raw contrast at 300 [mas] and \fwhm \,of the non-coronagraphic \psf, for a total of 111 different stars observed between 2016 and 2022 with IRDIS. For this, we processed a large number of individual frames that were obtained under different atmospheric conditions. We then compared the resulting raw contrast and the \psf \,shape as a function of the visible wave front sensor instant flux which scales with the G-band stellar magnitude. We repeated this analysis for the top 10$\%$ and 30$\%$ best turbulence conditions in Cerro Paranal.}
   {We found a strong decrease in the coronagraphic achievable contrast for star fainter than G $\sim$ 12.5 mag, even under the best atmospheric conditions. In this regime, the AO correction is dominated by the read-out noise of the \wfs \,detector. In particular we found roughly a factor ten decrease in the raw contrast ratio between stars with G $\sim$ 12.5 and G $\sim$ 14.0 mag. Similarly, we observe a sharp increase in the \fwhm \,of the non-coronagraphic \psf \,beyond G $\sim$ 12.5 mag, and a corresponding decrease in the strehl ratio from $\sim$ 50\% to $\sim$ 20\% for the faintest stars. Although these trend are observed for the two turbulence categories, the decrease in the contrast ratio and \psf \,sharpness is more pronounced for poorer conditions.  }
   {}


   \maketitle
%
%-------------------------------------------------------------------

\section{Introduction}

%- Include AN ELIPTICITY PLOT \newline
%- PLOT DISTRIBUTION OF SEEING AND TAU0 FOR GOOD CONTRAST CASES\newline
%- CHECK THE EFFECT OF THE SPATIAL FILTER (SMALLL, MEDIUM, LARGE) IN THE VWFS FLUX\newline
%MENTION CONTAMINATION IN THE CONTRAST BY DISKS. e.g. RU LUP, PDS70, EM_SR21!!!! \newline

%\textcolor{red}{Add SZ 51, ** CVS 155A,V V1012 Ori, 2MASS J12163970-7007355, TYC 8174-1586-1, GSC 08077-01788 } \newline
High-contrast direct imaging (DI) is a powerful technique for detecting and characterizing young and self-luminous giant planets, reflected light planets, highly polarized disks, Solar System objects, among other interesting objects. Thanks to its extreme Adaptive Optics (AO) system, called SAXO \citep{Fusco2006}, the Spectro-Polarimetric High-contrast Exoplanet REsearch instrument (SPHERE; \citealt{Beuzit2019}) is one of the most powerful instrument in the world for direct imaging science, and after more than 6 years of operations, it has led to the detection and characterization of a variety of (forming) planetary systems and protoplanet disks. 
Notable examples of these include $\beta$ Pictoris\,$b$ \citep{Lagrange2018}, 
the PDS\,70 planet forming system \citep{Muller2018}, the multiplanet systems around HD\,8799 \citep{Zurlo2019} and YSES 1 \citep{Bohn2020}, the TW Hydrae disk \citep{vanBoekel2017}, among others. \newline \indent
While different long-term SPHERE DI programs have mainly studied relatively bright stars (e.g. SHARDDS: \citealt{Milli2017a}; SHINE: \citealt{Desidera2021}; YSES: \citealt{Bohn2021}; BEAST: \citealt{Janson2021}), there is still a large population of stars located in young (age $\lesssim$ 10-20 Myr) stellar associations, which are ideal laboratories to study in-situ planet formation inside protoplanetary disks . However, these stars are embedded in gas and dust rich disks, which absorb a large fraction of the stellar visible light, making these object very red (see a discussion in \citealt{Boccaletti2020}). This effect is even stronger for M-type stars, which are intrinsically fainter in the optical when compared to the near infrared (nIR). 
The main disadvantage of targeting these kinds of stars with SPHERE is that the instant flux recorded in the visible wave front sensor (\wfs) is strongly reduced, and thus the AO correction is highly degraded. As more and more such faint stars (in visible light) are routinely being observed with SPHERE, it is mandatory to characterize the SAXO limiting magnitude. Motivated by this idea, we studied the SPHERE performance for targets fainter than G = 11.0 mag, where G corresponds to the Gaia Early Data Release 3 (Gaia EDR3) magnitude \citep{gaia_dr3}. \newline \indent
This paper is organized as follows: in section \ref{sec:data_analysis} we present the data used for the analysis we describe in details our methods. In section \ref{sec:results} we present the results of the raw contrast (from coronagraphic images) and the \fwhm \,of the non-coronagraphic \psf, as a function of the \wfs \,flux and the stellar G-band magnitude. Similarly, we computed the strehl ratio (\sr) for a handful of selected stars and we compare it with the \fwhm \,previously measured. In section \ref{sec:final_contrast} we compare the contrast from the raw images to that obtained in post-processing data.
Finally, the discussion and conclusions are presented in section \ref{sec:discussion}.

\section{Data and method \label{sec:data_analysis}}

\begin{table*}
\caption{Header keywords used to filter the data for the flux and coronagraphic observations. \label{tab:keywords}}
\centering
\begin{tabular}{lccc}
\hline\hline
\vspace{-0.3cm} \\
Keyword          &   Flux sequence                   & coronagraphic sequence   \\
\hline \vspace{-0.3cm} \\
DPR TYPE         &   OBJECT,FLUX                     &  OBJECT                  \\
DET SEQ1 DIT     &   $\geq$ 0.87 sec                 &  $\geq$ 2 sec            \\
DPR TECH         &   IMAGE / POLARIMETRY             &  IMAGE / POLARIMETRY    \\
%INS4 OPTI11 NAME &   B\_H                            &  B\_H                    \\ 
INS1 FILT NAME   &   B\_H                            &  B\_H                    \\ 
INS COMB IFLT    &   DB\_H23 /  DP$\_$0$\_$BB$\_$H   &  DB\_H23  /  DP$\_$0$\_$BB$\_$H  \\ 
AOS VISWFS MODE  &   GAIN$\_$1000$\_$FREQ$\_$300Hz   &  GAIN\_1000\_FREQ\_300Hz \\ 
INS4 FILT3 NAME  &   OPEN                            &  OPEN                    \\
INS4 OPTI22 ID   &   SMALL / MEDIUM / LARGE          & SMALL / MEDIUM / LARGE   \\
\hline 
\end{tabular}
\end{table*}

\begin{figure}[!h]
       \includegraphics[angle=0,scale=0.35]{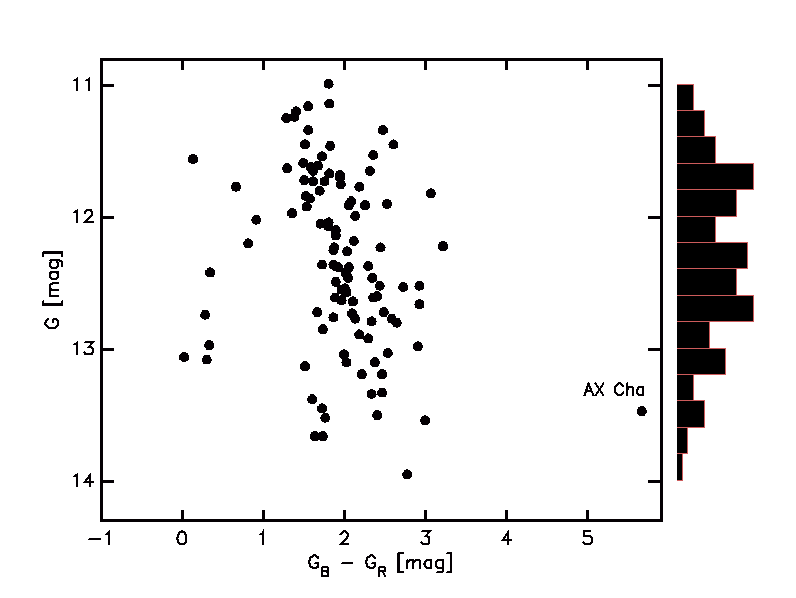}
    \caption{Gaia EDR3 color-magnitude diagram of all 111 stars included in the analysis. The typical uncertainties are smaller than the symbol size. The position of the reddest star in the sample (AX Cha) is also labeled. 
    The corresponding histogram of the G-band magnitude distribution is overplotted on the right side. 
    \label{color_magnitude}}
\end{figure}

\begin{figure}[!h]
       \includegraphics[angle=0,scale=0.35]{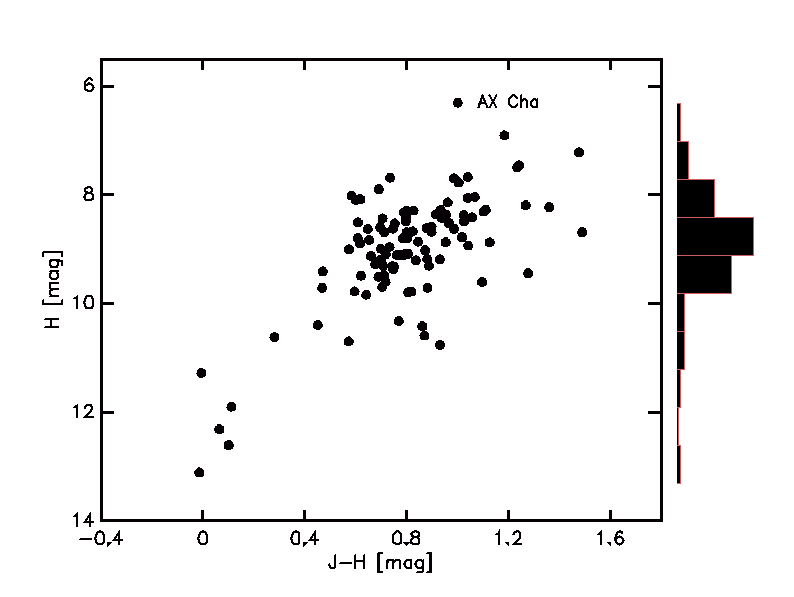}
    \caption{Same as Figure \ref{color_magnitude}, but for the J and H bands. \label{JH_col_mag}}
\end{figure}

We analyzed a total of 111 different stars observed with SPHERE between 2016 and 2022, all of them fainter than G = 11.0 mag, which are listed in Table \ref{stellar_mag}. 
Figure \ref{color_magnitude} shows the Gaia EDR3 color-magnitude diagram (CMD) with the position of all of the targets. Similarly, Figure \ref{JH_col_mag} shows a 2MASS All Sky Catalogue (\citealt{Cutri2003}) nIR CMD for our stars. 
The median G-band and H-band magnitude of the sample is 12.4 [mag] and 8.9 [mag], respectively.
We included in the analysis only H-band ($\lambda_{\rm c} \sim$ 1.63 $\mu$m) observations taken with the visible Wave Front Sensor (\texttt{WFS}) in the high gain (M = 1000) low frequency mode (300 Hz), with the \wfs \,spectral filter in the \texttt{OPEN} position. 
We also included both imaging and polarimetric observations. In the case of the coronagraphic observations we included only individual frames with detector integration time (\texttt{DIT}) longer than 2 seconds. The list of header keywords used to filter the data are summarized in Table \ref{tab:keywords}. 
We note that all of the raw data (non-coronagraphic FLUX and coronagraphic OBJECT data cubes) and their corresponding AO files (FITS binary tables containing atmospheric and SAXO telemetry data; see \citealt{Milli2017b}) were downloaded from the ESO archive. 
For the analysis, we developed an IDL-based code which computes the contrast for each individual raw frame, following the steps listed below. \newline \indent
First, we fitted a 2-D Gaussian function to each individual non-coronagraphic FLUX frames.
From the fit we estimated the peak flux ($f_{\texttt{nc}}$) and the corresponding \fwhm \,in the x- and y- pixels direction. The final $f_{\texttt{nc}}$ value was computed from the mean of all individual flux frames.  
We then used the mean $f_{\texttt{nc}}$ to predict the peak flux ($f_\texttt{c}$) in each individual coronagraphic OBJECT frames, by scaling the DIT, neutral density filters and other optical elements with different transmissions (e.g. when a sequence is obtained with and without the apodizer). We then estimated the noise level ($\sigma_{300}$) at 24 pixel from the coronagraph center (corresponding to $\sim$ 300 mas), by computing the standard deviation in a 5 pixels wide annulus. The resulting 5-$\sigma$ raw contrast at 300 mas (RC$_{300}$) was obtained from:

\begin{equation}
  RC_{300} = \frac{5\times \sigma_{300}}{f_{\texttt{c}}}
\end{equation}

We note that for broad band observations (B\_H filter) and differential polarimetry (DP$\_$0$\_$BB$\_$H filter) we used only the left side of the Infra-Red Dual-beam Imaging and Spectroscopy (IRDIS; \citealt{Dohlen2008}) detector, while for the dual-imaging mode (DB\_H23 filter) we analyzed independently the left ($\lambda_{\rm c} \sim$ 1.59 $\mu$m) and right ($\lambda_{\rm c} \sim$ 1.67 $\mu$m) side of the detector, and we computed then mean value from the two channels.
Finally, to study the effect of the atmospheric conditions on the contrast and the \psf, we retrieved the Differential Image Motion Monitor (DIMM) seeing and coherence time ($\tau_0$) at the beginning of each exposure, directly from the image header. 

\subsection{Instant flux on the visible \wfs}

As part of the SAXO system, SPHERE is equipped with a 40\,x\,40 Shack-Hartmann \wfs, (with a total of 1240 effective subapertures imaging the SPHERE entrance pupil). The light reaching the \wfs \,is recorded by a 240\,x\,240 pixels electron multiplying charge-coupled device (EMCCD) detector, sensitive to visible light ($\sim$ 0.45-0.95 $\mu$m). The main advantage of this detector is that it can operate at very high frequency (up to 1380 Hz), low read-out noise ($<$ 0.2 $e^{-}$/pixel/frame) and high multiplication gain (up to M = 1000; \citealt{Sauvage2016a}). However, as mentioned before, for our analysis we included only data taken with the EMCCD operating at 300Hz and M = 1000, which is the mode automatically selected by the real time computing system (SPARTA; \citealt{Fedrigo2006}) for faint targets (R $>$ 10.5 mag). \newline
The instant flux in the \wfs \,EMCCD detector (\fwfs) is recorded by SPARTA, and it is saved in the AO files. These files (among other telemetry data) contain the total flux per pupil, i.e., integrated in all 1240 subapertures. This flux (tagged \texttt{FLUX$\_$AVG} in the binary tables) is thus expressed in units of [ADU/frame/pupil]. To convert \texttt{FLUX$\_$AVG} into [e$^{-1}$/subaperture/frame], we used:

\begin{equation}
 f_{\scaleto{\texttt{WFS}}{3pt}} \,{\rm [e^{-1}/subaperture/frame] = \frac{g\times\texttt{FLUX\_AVG}}{n_s\times M}} ,
\end{equation}

where g = 16.5 [$e^-$/ADU], n${\rm _s}$\,=\,1240 [subapertures/pupil] and M = 1000 (unit-less).

\section{Results \label{sec:results}}

\subsection{Raw contrast and \fwhm \,as a function of the \wfs \,flux \label{sec:fwhm_fwfs}}

\begin{figure}[!h]
       \includegraphics[angle=0,scale=0.33]{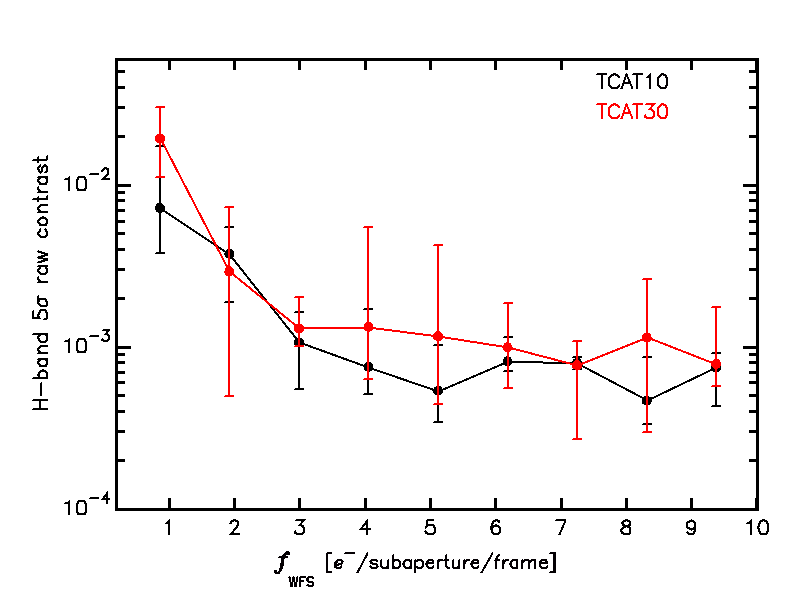}
    \caption{H-band 5$\sigma$ raw contrast at 300 [mas] as a function of \fwfs. The black and red lines correspond to atmospheric turbulence category \tcatone and \tcatthree, respectively.  \label{flux_rc}}
\end{figure}

\begin{figure}[!h]
       \includegraphics[angle=0,scale=0.33]{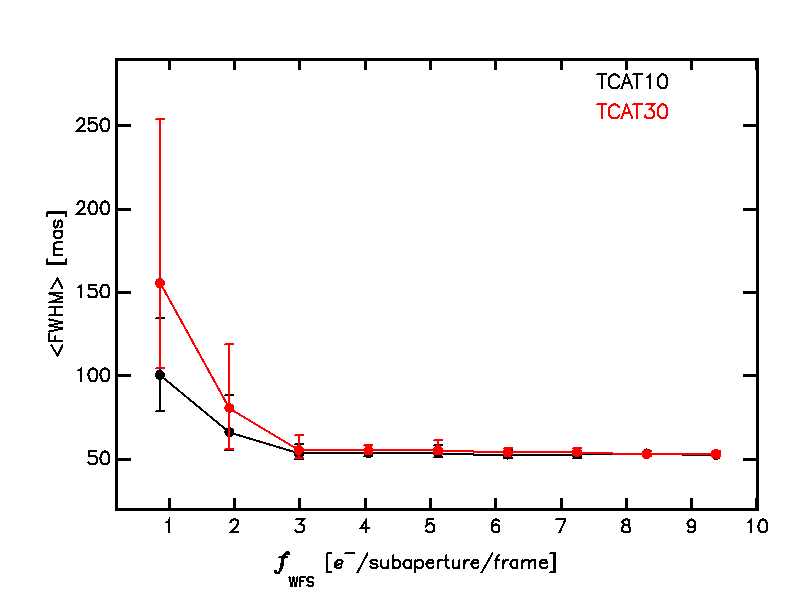}
    \caption{Mean full-width-at-half-maximum as a function of the visible \wfs \,flux. The black and red lines correspond to atmospheric turbulence category \tcatone and \tcatthree, respectively. \label{flux_fwhm}}
\end{figure}

After computing the raw contrast, we cleaned-up the sample by removing those frames leading to very deviant values. Typical examples of these include cases with the stars out of the coronagraph, AO loop open,  contamination from close binary companions, extreme cases of low-wind effect (\citealt{Sauvage2016b}; \citealt{Milli2018}), among others. We then compared the raw contrast as a function of \fwfs.
Figure \ref{flux_rc} shows the RC$_{300}$ versus the \fwfs binned in 1 [e$^{-}$/subaperture/frame] flux bins. We note that only frames with header values listed in Table \ref{tab:keywords} were included (as explained in section \ref{sec:data_analysis}).
The black dots correspond to atmospheric turbulence category \tcatone (seeing $<$ 0.6 arcsec; $\tau_0$ $>$ 5.2 ms) and the red dots to \tcatthree (seeing $<$ 0.8 arcsec; $\tau_0$ $>$ 4.1 ms). The error bars correspond to 1$\sigma$ equal-tailed confidence interval for each bin. As can be seen, there is a sharp decrease in the achievable contrast for \fwfs below $\sim$ 3 [e$^{-}$/subaperture/frame]. In this regime the AO correction is degraded due to the lack of photons collected by the \wfs. We also note that the \rct achieved under \tcatone conditions is slightly better than that under \tcatthree.
Similarly, Figure \ref{flux_fwhm} shows the mean \fwhm \,(from the x- and y- direction) measured on non-coronagraphic frames as a function of \fwfs. Again this quantity is computed on the left side of of the IRDIS detectors in the dual-band mode (see Table \ref{tab:keywords}).
It can be seen that for \fwfs level above $\sim$ 3 [e$^{-}$/subaperture/frame] the point-spread-function (\psf) 
%is close to the theoretical diffraction limit, with a 
is nearly flat with a \fwhm \,of $\sim$\,53 [mas], for both \tcatone and \tcatthree. On the other hand, for lower \fwfs values, the AO correction degrades and thus the \fwhm \,increases with decreasing flux level, reaching a mean value larger than $\sim$\,100 [mas], when reaching the sub-electron per frame regime, for both \tcatone and \tcatthree. 

\subsection{Correlation between stellar magnitude and \wfs \,flux}

\begin{figure}
       \includegraphics[angle=0,scale=0.33]{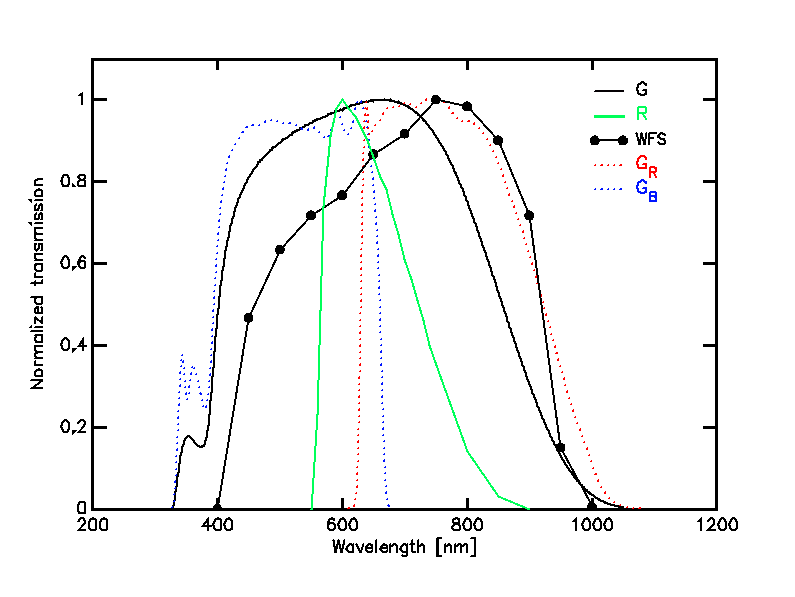}
    \caption{Normalized transmission curves for the Gaia EDR3 G (solid black line), G$_{\rm B}$ (dotted blue line), G$_{\rm R}$ (dotted red line) and Johnson R (solid green line) broad band filters. The total transmission curve (including upstream optics) of the \wfs \,is overplotted (solid black dots) \label{transmission}}
\end{figure}

\begin{figure}
       \includegraphics[angle=0,scale=0.33]{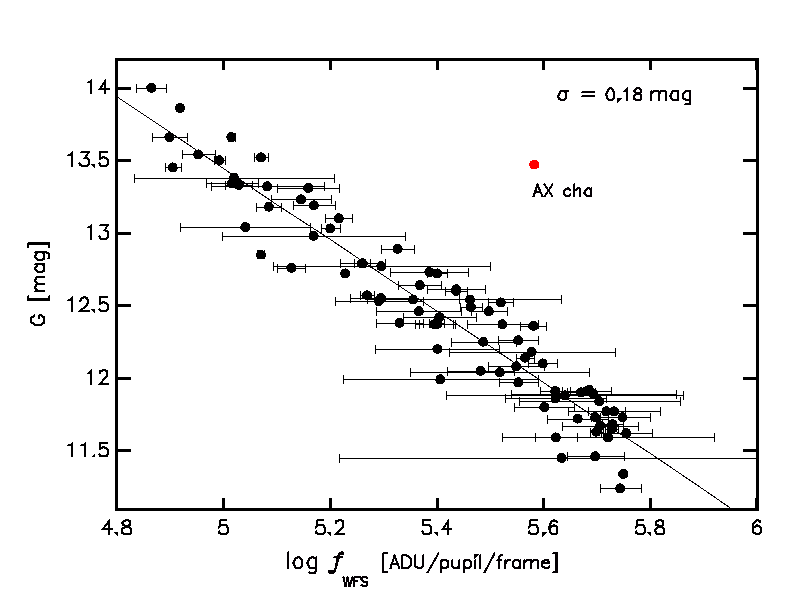}
    \caption{Stellar G-band magnitude versus the logarithm of the \wfs flux. The solid line corresponds to the best linear fit between these two quantities. The position of AX cha is also shown (red star). \label{logflux_gmag}}
\end{figure}

To understand the SPHERE performance not only as a function of \fwfs but also as a function of the target magnitude, we investigated the correlation between \fwfs and the stellar G-band magnitude. We choose this particular band because the sensitivity curve of the \wfs \,detector is very close to the G-band transmission curve and also since 
nearly all of the targets observed by SPHERE show a G-band entry in the EDR3 catalogue. 
Figure \ref{transmission} shows the normalized transmission of the Gaia EDR3 bands\footnote{https://www.cosmos.esa.int/web/gaia/edr3-passbands}, the Johnson R broad band filter \cite{Bessell1990} and the \wfs \,response, including all of the upstream optical elements and with the \wfs \,spectral filter in the OPEN position (which is the case for faint stars). 
It can be seen that the G-band transmission function is the closest one to the \wfs \,response function, although the latter one is slightly more sensitive to longer wavelength. 
For this reason, to predict the expected \fwfs value during an observing sequence of a faint target, the G-band should be used, instead of the R-band (as currently implemented in SPARTA). In fact, even though in general the value of these two magnitudes are very close, there are cases where the difference is as large as one magnitude (corresponding to a factor $\sim$ 2.5 in flux). 
For this analysis, we first corrected the flux in the \wfs \,for the target airmass, using an extinction coefficient of 0.12 \citep{Patat2011}. We then compared this value with the stellar G-band magnitude. We note that to minimize the stellar color effect, we only included stars with 1 $<$ G$_B$ - G$_R$ $<$ 3 (see Figure \ref{color_magnitude}). The results are shown in Figure \ref{logflux_gmag}. The error bars in the \wfs \,flux correspond to the standard deviation from all individual measurements. The solid line corresponds to the \wfs \,magnitude scale ($m_{\scaleto{\texttt{WFS}}{3pt}}$), given by:

\begin{equation}
  m_{\scaleto{\texttt{WFS}}{3pt}} = -2.5\, log(f_{\scaleto{\texttt{WFS}}{3pt}}) + 25.7
\end{equation}

\noindent We note that we removed outliers beyond 2.5\,$\sigma$ from the fit, leading to a scatter of 0.18 mag. This value is mainly explained by observations performed under different observing conditions (sky transparency, seeing, etc) and more importantly due to the fact that we included observations using different spatial filters \citep{Fusco2016}, as listed in Table \ref{tab:keywords}, meaning that a different amount of the light is blocked for different observations. Ideally, this analysis should be repeated with data taken under photometric conditions and using the \wfs \,spatial filter in the OPEN position, as opposed to the majority of the observations analyzed here that were done using either the SMALL or MEDIUM spatial filter.

\begin{figure}
       \includegraphics[angle=0,scale=0.33]{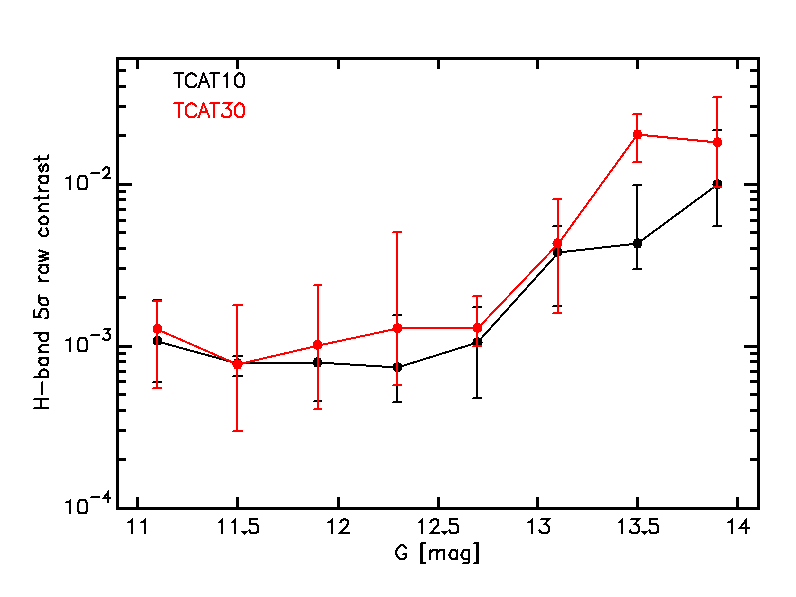}
    \caption{5-$\sigma$ raw contrast at 300 mas as a function of the stellar G-band magnitude. The black and red lines correspond to atmospheric turbulence category \tcatone and TCAT30, respectively. \label{fig:gmag_rc}}
\end{figure}

\begin{figure}[!h]
       \includegraphics[angle=0,scale=0.34]{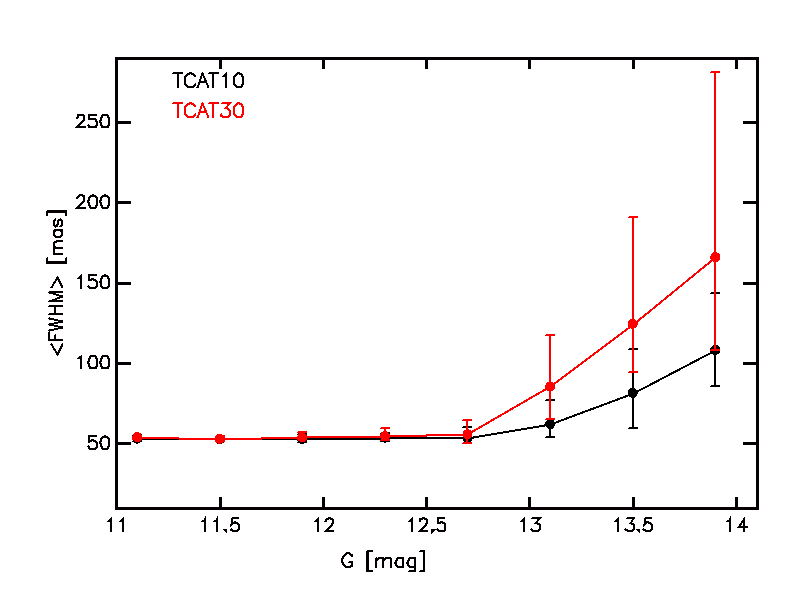}
    \caption{Mean full-width-at-half-maximum as a function of the stellar G-band magnitude. The solid black and red dots (lines) correspond to atmospheric turbulence category \tcatone and TCAT30, respectively. 
    \label{Gband_fwhm}}
\end{figure}

Using this correlation we can directly compare the stellar G magnitude and the expected raw contrast, for different atmospheric conditions. We therefore replicated Figure \ref{flux_rc} and Figure \ref{flux_fwhm}, but this time we replaced \fwfs by the stellar G magnitude. The results are shown in Figures \ref{fig:gmag_rc} and \ref{Gband_fwhm}. As can be seen, there is roughly one order of magnitude decrease in the contrast ratio beyond G $\sim$ 12.5. Similarly, there is an increase of the \fwhm \,of the \psf \,from $\sim$ 53 [mas] to  $\gtrsim$ 100 [mas] between G $\sim$ 12.5 [mag] and $\sim$ 14 [mag]. 

\subsection{Strehl ratio \label{sec:strehl1}}

\begin{figure}[!h]
       \includegraphics[angle=0,scale=0.40]{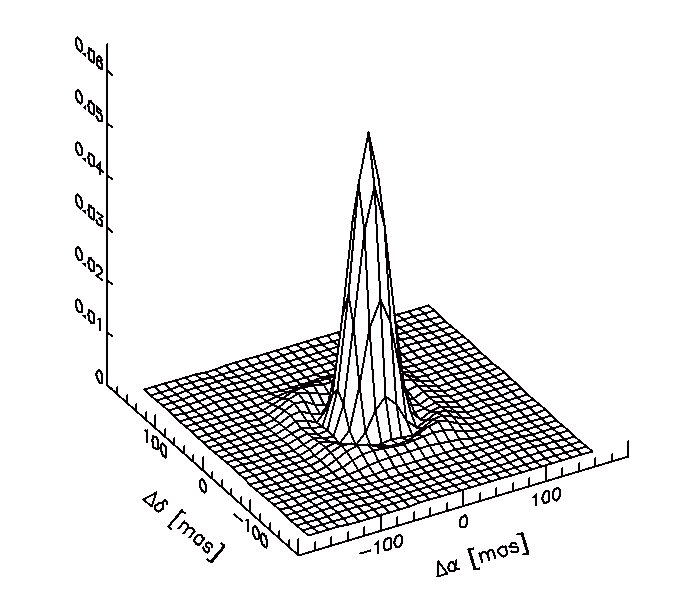}
    \caption{Normalized diffraction limited theoretical \psf \,of the VLT pupil at 1.63 $\mu$m, as seen by IRDIS/SPHERE. \label{fig:theoretical_psf}}
\end{figure}

\begin{figure*}[!h]
       \includegraphics[angle=0,scale=0.40]{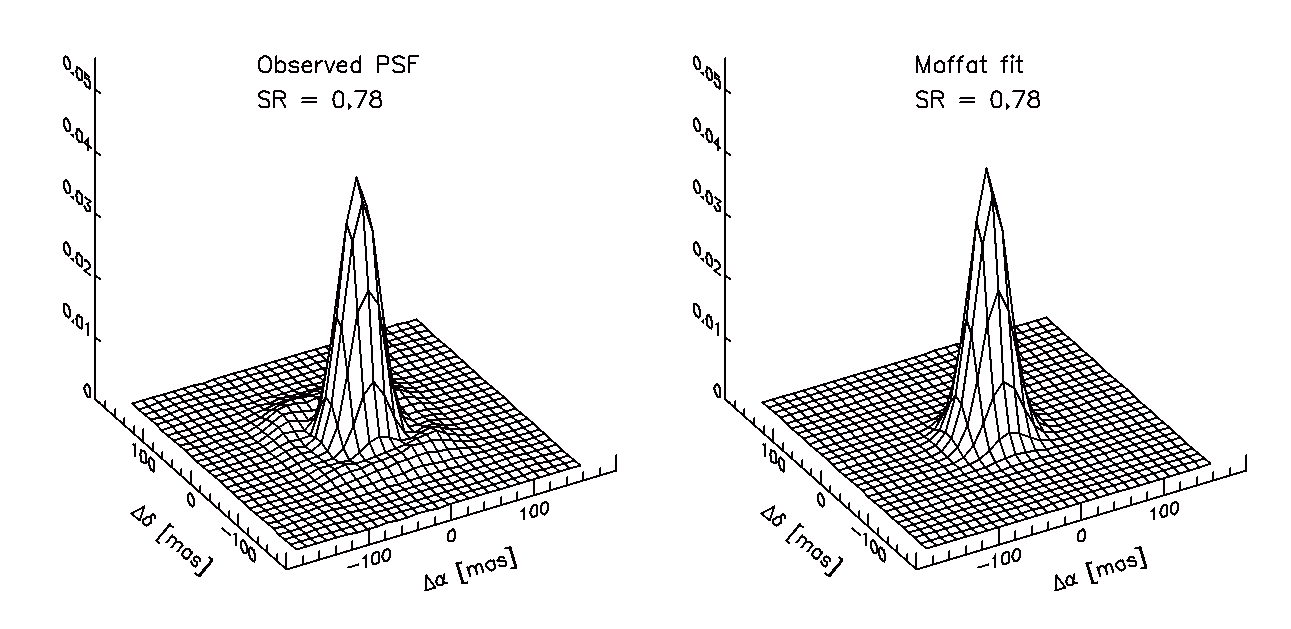}
    \caption{Observed \psf \,of a star leading to a strehl ratio of 0.78 (left) and its corresponding  Moffat fit (right). \label{fig:observed_psf}}
\end{figure*}

\begin{figure}[!h]
       \includegraphics[angle=0,scale=0.34]{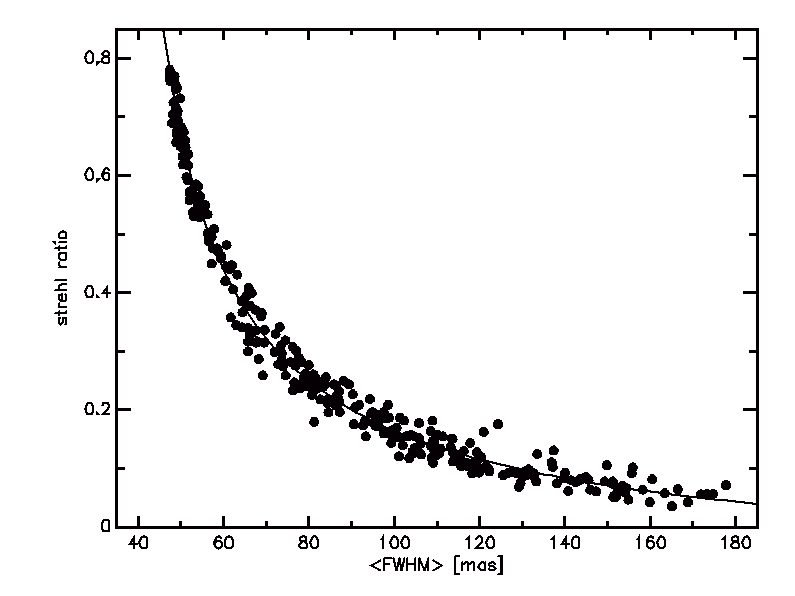}
    \caption{Strehl ratio versus the \fwhm \,measured from FLUX frame. The solid line corresponds to the fit following equation \ref{eq:fwhm_sr_fit} \label{fig:fwhm_sr}}
\end{figure}

In addition to the \fwhm \,values, we investigated the dependence of the strehl ratio (SR) with the \fwfs and the stellar magnitude. 
For this, we first computed a theoretical diffraction limited \psf \,for the VLT as seen by SPHERE, (i.e, including the effect of the Lyot stop and the apodizer in the optical path) at 1.63 $\mu$m . The \psf \,was computed over an area of 29x29 pixels (including light up to the third airy ring) with the IRDIS plate scale of 12.27 [mas/pix], and was then normalized by the total integrated flux in all 29x29 pixels area. A 3D view of the resulting normalized \psf \,is shown in Figure \ref{fig:theoretical_psf}. 
We then fitted a two-dimensional Moffat profile \citep{Moffat1969} of the form:
\begin{equation}
  I(x,y) = I_0 / (U+1)^\beta, \label{eq:moffat_fit}
\end{equation}
where $U$ corresponds to:
\begin{equation}
  U = [(x-x_0)/A_x]^2 + [(y-y_0)/A_y]^2 
  \label{eq:moffat_U}
\end{equation}
The resulting coefficients are listed in Table \ref{tab:coef_moffat_psf}. 
We then repeated the Moffat fit, but this time to a total of 332 observed FLUX frames. We note that in this case we included also brighter stars (leading to higher SR) that were not part of the sample presented in section \ref{sec:data_analysis}. 
%Also, we did not include a zero point to account for the background, since in all of the FLUX frames analyzed this value is negligible ($\lesssim$ 2-3 [ADU/pixel]) due to the use of ND filters and typical DIT = 0.87 sec. 
The resulting SR was simply estimated by taking the ratio between the peak of the Moffat fit (corresponding to I$_0$ in eq. \ref{eq:moffat_fit}) of the observed FLUX frames and that from the theoretical \psf. 
An example of an observed \psf \,leading to \sr \,= 0.78 and its corresponding Moffat fit is shown in Figure \ref{fig:observed_psf}.   \newline \indent
\begin{table}
\caption{Resulting coefficients for the Moffat fit of the theoretical \psf \,at 1.63 $\mu$m, following equations \ref{eq:moffat_fit} and \ref{eq:moffat_U}. \label{tab:coef_moffat_psf}}
\centering
\begin{tabular}{lr}
\hline\hline
\vspace{-0.3cm} \\
Coefficient         &   Value  \\
\hline \vspace{-0.3cm} \\
$I_0$  &    0.058  \\
$\beta$  &  26710.9   \\
$A_x$  &   347.5\\
$A_y$  &  345.7 \\
$x_0$  &  14.0 \\
$y_0$  &  14.0 \\
\hline 
\end{tabular}
\end{table}
In addition, we measured the \fwhm \,to the FLUX frames (as explained in section \ref{sec:fwhm_fwfs}), and we compared them with the measured \sr. The results are presented in Figure \ref{fig:fwhm_sr}. We finally obtained a correlation between the \sr \,and \fwhm \,by fitting a function of the form:

\begin{equation}
\label{eq:fwhm_sr_fit}
% \sr = a_0 + a_1\times\fwhm +a_2\times\fwhm^2 + a_3\times\fwhm^3.
  \sr = a + b/(\fwhm+c)^\alpha 
\end{equation}

The resulting coefficients are listed in Table \ref{tab:coef}. The RMS of the post-fit residuals is 0.023. The best-fit is also over plotted in Figure \ref{fig:fwhm_sr}

\begin{table}
\caption{Resulting coefficients for the \sr-\fwhm \,correlation presented in equation \ref{eq:fwhm_sr_fit}. \label{tab:coef}}
\centering
\begin{tabular}{lr}
\hline\hline
\vspace{-0.3cm} \\
Coefficient         &   Value  \\
\hline \vspace{-0.3cm} \\
$a$  &    -0.099  \\
$b$  &   5.296  \\
$c$  &   -33.236 \\
$\alpha$  &  0.716 \\
\hline 
\end{tabular}
\end{table}

\begin{figure}[!h]
       \includegraphics[angle=0,scale=0.34]{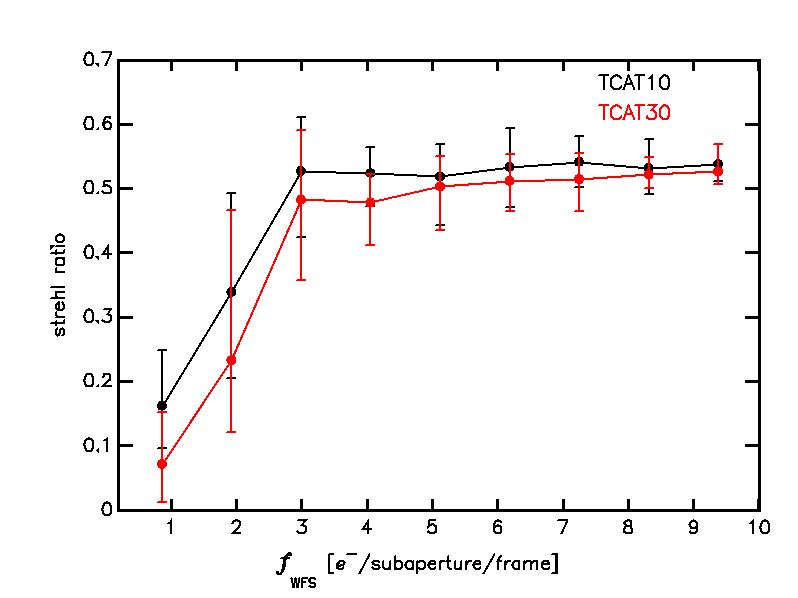}
    \caption{Strehl ratio as a function of the visible \wfs \,flux. The black and red lines correspond to atmospheric turbulence category \tcatone and \tcatthree, respectively. \label{fig:fwfs_sr}}
\end{figure}

\begin{figure}[!h]
       \includegraphics[angle=0,scale=0.33]{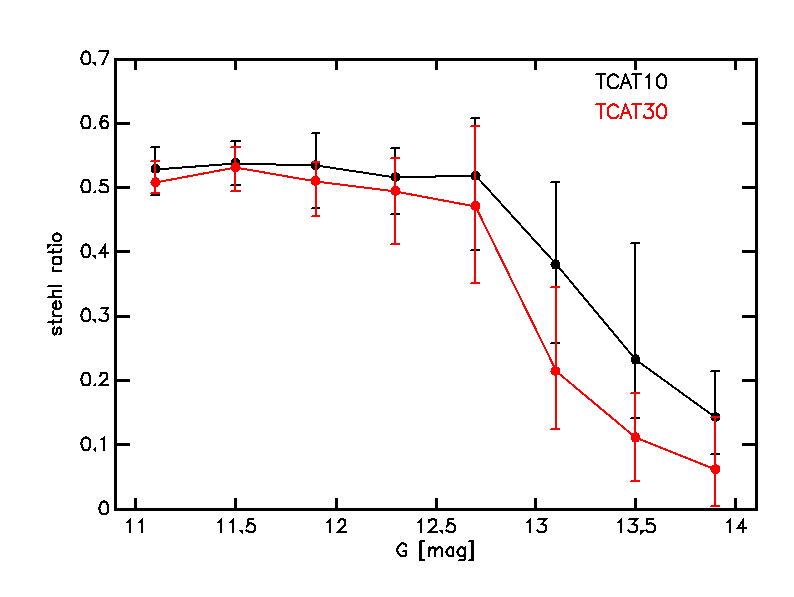}
    \caption{Strehl ratio as a function of the stellar G-band magnitude. The solid black and red dots (lines) correspond to atmospheric turbulence category \tcatone and \tcatthree, respectively.  \label{Gband_strehlratio}}
\end{figure}

Using this correlation, we then converted the \fwhm \,values into \sr. \,We therefore replicated Figures \ref{flux_fwhm} and \ref{Gband_fwhm}, but this time we replaced the \fwhm \,by the \sr. The results are presented in Figures \ref{fig:fwfs_sr} and \ref{Gband_strehlratio}.

\section{Final contrast \label{sec:final_contrast}}

\begin{figure}[!h]
\includegraphics[angle=0,scale=0.33]{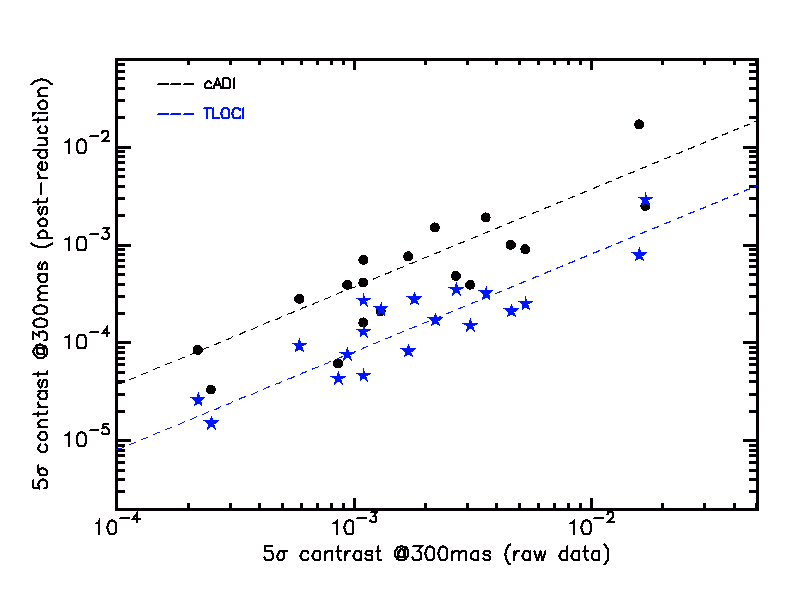}
\caption{Comparison between the contrast measured at 300 [mas] from individual raw frames and post-reduction full observing sequence. The black dots and blue stars correspond to cADI and TLOCI reduction techniques, respectively. \label{Fig:raw2red300}}
\end{figure}

To investigate what is the achievable final contrast as a function of the stellar magnitude, we compared the raw contrast with that measured after post-processing. We included in this analysis 19 different stars, leading to \rct values covering roughly two orders of magnitude (between $\sim$ 2$\times$10$^{-2}$ and 2$\times$10$^{-4}$).
For this we retrieved the post-processing contrast curves (corrected for the throughput of the algorithm) produced by the SPHERE Data Center \citep{delorme2017}, for the classical Angular Differential Imaging (cADI; \citealt{Marois2006}) and the Template Local Optimized Combination of Images (TLOCI; \citealt{Marois2010}; \citealt{Galicher2018}) techniques. 
We note that we only included relatively long observing sequences, leading to a total field of view (FoV) rotation $>$ 15 degrees, to avoid strong self-subtraction and thus low throughput. These results are presented in Figure \ref{Fig:raw2red300}. The black dots and blue stars correspond to cADI and TLOCI reduction techniques. The dashed black and blue lines correspond to the median conversion factor, corresponding to 2.7 and 12.4 for the cADI and TLOCI techniques respectively. We can then use this result to estimate the final contrast on post-processing data by scaling \rct using these two values (e.g. to convert Figures \ref{flux_rc} and \ref{fig:gmag_rc} into final contrast). Finally, we repeated this conversion, but this time from \rct to final contrast at 500 [mas]. We obtained median conversion factors of 11.0 and 43.5 for the cADI and TLOCI techniques, respectively. \newline

\section{Summary and discussion \label{sec:discussion}}

In this paper we have analyzed a sample of 111 relatively faint stars (G $\ge$ 11.0 mag) observed with IRDIS/SPHERE between 2016 and 2021, with the aim of investigating how the AO correction (and thus the final performance in terms of achievable contrast) degrades when approaching the limiting magnitude regime. 
For this we measured the H-band \rct from coronagraphic frames and we compared this quantity to the instant flux received by the visible \wfs. We found that the \rct is at the $\sim$\,10$^{-3}$ level for \fwfs $\gtrsim$ \,3 [e$^{-1}$/subaperture/frame]. 
For \fwfs below this value, the AO correction strongly degrades, decreasing the \rct by a factor $\sim$ 10 when reaching the 1  [e$^{-1}$/subaperture/frame] level. In terms of stellar magnitude, this turning point is observed at G $\sim$ 12.5 [mag]. At G $\sim$ 14 [mag] the measured \rct is at the 10$^{-2}$ level.  We observed this behaviour in the data taken under both \tcatone and \tcatthree atmospheric conditions, with the former one performing slightly better. \newline \indent
Similarly, we computed the \fwhm \,and \sr \,from the FLUX frames. We observed a very similar trend above and below the $\sim$ 3 [e$^{-1}$/subaperture/frame] limit. In particular, we measured a factor $\sim$ 2 and 3 increase in the \fwhm \,from this limit and the $\sim$ 1 [e$^{-1}$/subaperture/frame] \fwfs level, for \tcatone and \tcatthree, respectively. Similarly, the \sr \,drops from above 0.5 to less than 0.2, for both \tcatone and \tcatthree. \newline \indent
Finally, we computed the final contrast at 300 [mas] and 500 [mas], in a selected sample of stars, using the post-reduction contrast curves. We obtained a median conversion factors at 300 [mas] of $\sim$ 3 and 12, for the cADI and TLOCI techniques, respectively. 
Similarly, we found scaling factors to 500 [mas] of $\sim$ 11 and 44 for cADI and TLOCI, respectively. Using these conversion factors, we can then estimate the final contrast achievable at 300 [mas] and 500 [mas], by scaling the results presented in this work. \newline \indent
As an example, let us consider a star with the exact same luminosity and colors of the well known planet host star PDS70 \citep{Keppler2018}. Let us also consider that such star is located at $\sim$ 213 [pc] from the Earth, so that it has: G = 13.0 [mag] and  H = 11.2 [mag]. Following Figure \ref{fig:gmag_rc}, we can expect a  
\rct of $\sim$ 4$\times$10$^{-3}$ under good atmospheric conditions (\tcatthree or better). If we perform a relatively long (FoV rotation $\gtrsim$ 15 deg) ADI sequence, we can then estimate a 5$\sigma$ final contrast at 500 [mas] using the TLOCI technique of $\sim$ 10$^{-4}$. If we compare these number with the  
AMES-COND models \citep{Allard2001}, this limit corresponds to a 10 [Myr] planet with a $\sim$ 4.8 \mjup mass and located at a physical separation of $\sim$ 107 [AU] from the host star. \newline \indent
Improving the contrast on faint targets and reaching fainter or redder stars represent one of the primary scientific drivers for the instrument upgrade called SPHERE+ \citep{Boccaletti2020}. Indeed ALMA surveys such as DSHARP \citep{Andrews2018} revealed dozens of star-forming disks with structures such as gaps, rings or spirals, suggesting planet formation is on-going. PDS70 represents one of the brightest examples of this population, the bulk of it being currently out of reach for SPHERE (see Fig. 12 of \citealt{Boccaletti2020}). Coupling the current AO correction stage to a more sensitive second-stage AO in cascade represents the main technical improvement of SPHERE+ \citep{2Cerpa-Urra2022} and the characterisation of the current performance on faint targets presented here is crucial to make informed technical choices in the design of the second stage AO for SPHERE+.

\begin{acknowledgements}
This work has made use of the the SPHERE Data Centre, jointly operated by OSUG/IPAG (Grenoble), PYTHEAS/LAM/CESAM (Marseille), OCA/Lagrange (Nice), Observatoire de Paris/LESIA (Paris), and Observatoire de Lyon. 
This research has made use of the SIMBAD database, operated at CDS, Strasbourg, France. 
 This work has made use of data from the European Space Agency (ESA) mission
 {\it Gaia} (\url{https://www.cosmos.esa.int/gaia}), processed by the {\it Gaia}
 Data Processing and Analysis Consortium (DPAC,
 \url{https://www.cosmos.esa.int/web/gaia/dpac/consortium}). Funding for the DPAC
 has been provided by national institutions, in particular the institutions
 participating in the \gaia Multilateral Agreement.

\end{acknowledgements}

\bibliographystyle{aa}
\bibliography{sphere_bib}
%\bibliography{sphere}
%\begin{thebibliography}{}
%\end{thebibliography}

\begin{appendix}
\onecolumn
\section{Targets apparent magnitude}
\begin{longtable}{l c c c c c r}
\caption{Gaia EDR3 and 2MASS All Sky Catalogue $JHK$ stellar magnitudes, listed in ascending order. 
\label{stellar_mag} }\\
%\centering
%\begin{tabular}{lcccccr}
\hline \hline 
\vspace{-0.3cm} \\
G     &  G$_B$ &  G$_R$  &  $J$    &  $H$    &  $K$   &                Star name \\
\hline \vspace{-0.3cm}  \\
\endhead
\hline
\multicolumn{7}{r}{\footnotesize\itshape Continue on the next page}
\endfoot
\hline
\endlastfoot
  10.99 &   11.81 &   10.00 &    8.4 &    7.7 &    7.2 & 2MASS-J16274028-2422040 \\
  11.08 &   11.51 &   10.43 &    9.7 &    9.2 &    9.1 & 2MASS-J13015435-4249422 \\
  11.14 &   11.88 &   10.06 &    8.7 &    8.1 &    7.9 &          TYC-8083-455-1 \\
  11.16 &   11.88 &   10.32 &    9.4 &    8.7 &    8.5 & 2MASS-J10560422-6152054 \\
  11.20 &   11.83 &   10.42 &    9.3 &    8.5 &    8.0 & 2MASS-J11122441-7637064 \\
  11.24 &   11.86 &   10.47 &    9.5 &    8.9 &    8.8 & 2MASS-J12383556-5916438 \\
  11.25 &   11.68 &   10.39 &    9.4 &    8.8 &    8.7 & 2MASS-J13121764-5508258 \\
  11.34 &   12.67 &   10.19 &  \dots &  \dots &  \dots &                HIP20990 \\
  11.34 &   12.06 &   10.50 &    8.6 &    7.9 &    7.7 & 2MASS-J00172353-6645124 \\
  11.45 &   12.14 &   10.62 &    8.6 &    8.0 &    7.8 & 2MASS-J12072738-3247002 \\
  11.45 &   12.88 &   10.27 &    9.4 &    8.6 &    8.2 & 2MASS-J11175186-6402056 \\
  11.46 &   12.34 &   10.51 &    9.1 &    8.3 &    8.0 &                  DN-Tau \\
  11.53 &   12.90 &   10.54 &    8.7 &    7.7 &    7.1 &                  HQ-Tau \\
  11.54 &   12.35 &   10.62 &    9.1 &    8.3 &    7.8 &                 CHX-18N \\
  11.56 &   11.60 &   11.46 &   11.3 &   11.3 &   11.3 & 2MASS-J21012926-0622148 \\
  11.59 &   12.27 &   10.77 &    9.6 &    9.0 &    8.9 & 2MASS-J16015918-3612555 \\
  11.61 &   12.39 &   10.71 &    9.6 &    8.8 &    8.5 &              CD-40-8434 \\
  11.62 &   12.36 &   10.77 &    9.7 &    9.0 &    8.8 & 2MASS-J12510556-5253121 \\
  11.63 &   12.20 &   10.90 &    9.9 &    9.4 &    9.3 & 2MASS-J12404664-5211046 \\
  11.65 &   12.87 &   10.55 &    9.1 &    8.4 &    8.2 & 2MASS-J22025453-6440441 \\
  11.65 &   12.35 &   10.73 &    8.8 &    7.8 &    6.9 & 2MASS-J04470620+1658428 \\
  11.67 &   12.55 &   10.73 &    9.3 &    8.6 &    8.4 & 2MASS-J04331003+2433433 \\
  11.68 &   12.65 &   10.70 &    9.5 &    8.7 &    8.6 & 2MASS-J04440099-6624036 \\
  11.72 &   12.41 &   10.90 &    9.5 &    8.7 &    8.2 & 2MASS-J18425797-3532427 \\
  11.73 &   12.41 &   10.79 &    9.3 &    8.6 &    8.3 & 2MASS-J04551098+3021595 \\
  11.73 &   12.47 &   10.71 &    9.3 &    8.4 &    7.8 &                  SZ-cha \\
  11.75 &   12.71 &   10.75 &    9.2 &    8.4 &    8.0 & 2MASS-J15392776-3446171 \\
  11.77 &   12.02 &   11.35 &   10.9 &   10.6 &   10.5 & 2MASS-J00190738-3712352 \\
  11.77 &   12.90 &   10.71 &    9.1 &    8.3 &    8.0 & 2MASS-J16303563-2434188 \\
  11.80 &   12.59 &   10.89 &    9.7 &    8.9 &    8.7 & 2MASS-J11040909-7627193 \\
  11.82 &   13.61 &   10.54 &    8.7 &    8.1 &    7.8 &                  TX-PsA \\
  11.84 &   12.54 &   11.01 &   10.0 &    9.3 &    9.2 &         RX-J1555.6-3709 \\
  11.86 &   12.59 &   11.01 &   10.0 &    9.3 &    9.2 &         GSC-06214-00210 \\
  11.88 &   12.86 &   10.77 &    8.7 &    7.7 &    7.0 &                  DG-Tau \\
  11.90 &   13.26 &   10.73 &    9.1 &    8.5 &    8.3 & 2MASS-J02014693+0117059 \\
  11.91 &   13.07 &   10.81 &    9.2 &    8.3 &    7.8 &                  DE-Tau \\
  11.91 &   12.94 &   10.88 &    9.5 &    8.8 &    8.6 & 2MASS-J08422372-7904030 \\
  11.92 &   12.39 &   10.85 &    9.3 &    8.5 &    8.1 & 2MASS-J02581122+2030037 \\
  11.97 &   12.57 &   11.21 &   10.2 &    9.7 &    9.5 &         GSC-08584-01898 \\
  11.99 &   13.03 &   10.89 &    9.1 &    8.1 &    7.5 &                  GK-Tau \\
  12.02 &   12.39 &   11.47 &   10.8 &   10.4 &   10.3 & 2MASS-J06281844-7621467 \\
  12.04 &   12.92 &   11.11 &    9.8 &    9.1 &    8.9 & 2MASS-J16093030-2104589 \\
  12.05 &   12.81 &   11.10 &    9.5 &    8.6 &    8.0 &                  DS-Tau \\
  12.07 &   12.93 &   11.12 &    9.6 &    8.8 &    8.4 & 2MASS-J16035767-2031055 \\
  12.10 &   13.02 &   11.12 &    9.7 &    9.0 &    8.9 & 2MASS-J16071370-3839238 \\
  12.14 &   13.01 &   11.11 &    9.6 &    8.7 &    8.0 &                  DL-Tau \\
  12.18 &   13.23 &   11.11 &    9.5 &    8.4 &    7.8 &                  CI-Tau \\
  12.20 &   12.50 &   11.68 &   10.6 &    9.7 &    8.8 & 2MASS-J05113654-0222484 \\
  12.22 &   14.11 &   10.89 &    8.1 &    6.9 &    6.2 &                Haro-1-6 \\
  12.23 &   13.09 &   11.21 &    9.5 &    8.7 &    8.0 &            V*-V1279-Sco \\
  12.23 &   13.51 &   11.06 &    8.7 &    7.2 &    6.1 &                  WW-Cha \\
  12.25 &   13.06 &   11.19 &    9.4 &    8.3 &    7.9 & 2MASS-J16153456-2242421 \\
  12.26 &   13.28 &   11.24 &    9.5 &    8.6 &    8.1 & 2MASS-J16221852-2321480 \\
  12.36 &   13.21 &   11.48 &   10.1 &    9.3 &    8.7 & 2MASS-J05383368-0244141 \\
  12.36 &   13.26 &   11.39 &    9.9 &    9.1 &    8.8 & 2MASS-J13042410-7650012 \\
  12.37 &   13.28 &   11.39 &    9.9 &    9.1 &    8.6 & 2MASS-J15591647-4157102 \\
  12.37 &   13.56 &   11.26 &  \dots &  \dots &  \dots &                  UZ-Tau \\
  12.37 &   13.28 &   11.39 &    9.9 &    9.1 &    8.6 &            WRAY-15-1400 \\
  12.38 &   13.34 &   11.28 &    9.5 &    8.2 &    7.3 &                  DO-Tau \\
  12.38 &   13.33 &   11.40 &    9.8 &    9.1 &    8.5 & 2MASS-J10180498-5816263 \\
  12.42 &   13.43 &   11.41 &    9.8 &    8.9 &    8.3 &                  IP-Tau \\
  12.42 &   12.54 &   12.19 &   12.0 &   11.9 &   11.8 &         TYC-9337-2511-1 \\
  12.46 &   13.69 &   11.34 &    9.5 &    8.5 &    8.0 &                  DQ-Tau \\
  12.46 &   13.47 &   11.42 &    9.8 &    8.8 &    8.2 &                  DH-Tau \\
  12.49 &   13.41 &   11.51 &   10.1 &    9.3 &    9.0 & 2MASS-J16100501-2132318 \\
  12.52 &   13.50 &   11.06 &    9.1 &    8.1 &    7.5 &                Haro-1-4 \\
  12.52 &   14.20 &   11.27 &    8.7 &    7.5 &    6.9 & 2MASS-J16283266-2422449 \\
  12.53 &   14.00 &   11.27 &   10.7 &    9.6 &    9.0 &                V409-Tau \\
  12.54 &   13.54 &   11.53 &    9.9 &    9.0 &    8.7 &               V1094-Sco \\
  12.55 &   13.53 &   11.56 &   10.0 &    9.2 &    8.5 &                  PDS-65 \\
  12.57 &   13.58 &   11.55 &   10.2 &    9.5 &    9.3 & 2MASS-J13342461-6517473 \\
  12.60 &   13.88 &   11.47 &    9.9 &    9.1 &    8.8 &                  CX-Tau \\
  12.61 &   13.54 &   11.65 &   10.4 &    9.7 &    9.6 & 2MASS-J15545286-3827566 \\
  12.61 &   13.87 &   11.51 &    9.3 &    8.4 &    7.9 &                  GI-Tau \\
  12.63 &   13.58 &   11.61 &   10.1 &    9.4 &    9.2 & 2MASS-J16064385-1908056 \\
  12.64 &   13.71 &   11.60 &    9.9 &    9.1 &    8.6 &                V836-Tau \\
  12.66 &   14.33 &   11.40 &    8.7 &    7.5 &    6.7 & 2MASS-J16271027-2419127 \\
  12.72 &   13.40 &   11.73 &    9.9 &    9.1 &    8.5 & 2MASS-J05065551-0321132 \\
  12.72 &   14.06 &   11.57 &    9.9 &    9.2 &    8.9 &                  FP-Tau \\
  12.73 &   13.80 &   11.70 &   10.1 &    9.5 &    9.3 & 2MASS-J18443116-3723347 \\
  12.74 &   12.84 &   12.55 &   12.4 &   12.3 &   12.3 &             WD-1202-232 \\
  12.76 &   13.67 &   11.80 &   10.1 &    9.2 &    8.6 &                  TW-Cha \\
  12.77 &   14.15 &   11.56 &    9.4 &    8.4 &    7.8 & 2MASS-J16262367-2443138 \\
  12.77 &   13.69 &   11.55 &    9.6 &    8.2 &    7.1 &                  CW-Tau \\
  12.79 &   14.02 &   11.68 &    9.6 &    8.6 &    7.6 &                 PDS-198 \\
  12.80 &   14.26 &   11.61 &    9.4 &    8.3 &    7.9 & 2MASS-J16244839-2359160 \\
  12.85 &   13.70 &   11.96 &   10.6 &    9.8 &    9.4 & 2MASS-J05401286-0222020 \\
  12.89 &   14.02 &   11.83 &   10.2 &    9.5 &    9.2 & 2MASS-J16090075-1908526 \\
  12.92 &   14.12 &   11.82 &   10.2 &    9.5 &    9.2 & 2MASS-J16111534-1757214 \\
  12.97 &   13.09 &   12.75 &   12.7 &   12.6 &   12.6 &                 LAWD-25 \\
  12.98 &   14.64 &   11.73 &  \dots &  \dots &  \dots &                V710-Tau \\
  13.03 &   14.32 &   11.78 &    9.5 &    8.5 &    7.6 &                  HP-Tau \\
  13.04 &   14.02 &   12.02 &   10.3 &    9.6 &    9.2 &          Ass-Cha-T-1-32 \\
  13.06 &   13.08 &   13.05 &   13.1 &   13.1 &   12.7 & 2MASS-J23284760+0514540 \\
  13.08 &   13.19 &   12.88 &  \dots &  \dots &  \dots &   CRTS-J232812.6-395523 \\
  13.10 &   14.07 &   12.04 &   10.4 &    9.8 &    9.5 &                  DM-Tau \\
  13.10 &   14.34 &   11.96 &   10.1 &    9.2 &    8.6 & 2MASS-J15464473-3430354 \\
  13.13 &   13.84 &   12.32 &   11.3 &   10.7 &   10.6 & 2MASS-J06000075-3100276 \\
  13.19 &   14.28 &   12.06 &   10.0 &    8.9 &    8.4 & 2MASS-J16154416-1921171 \\
  13.19 &   14.49 &   12.02 &    9.4 &    8.4 &    7.8 &                  IQ-Tau \\
 % 13.23 &   14.45 &   11.54 &    9.7 &    9.2 &    8.9 & 2MASS-J10140807-7636327 \\
  13.33 &   14.65 &   12.18 &   10.2 &    9.3 &    8.9 & 2MASS-J13045571-7739495 \\
  13.34 &   14.51 &   12.17 &   10.5 &    9.8 &    9.4 &                  ET-Cha \\
  13.38 &   14.10 &   12.49 &   10.7 &    9.5 &    8.4 &                  HN-Tau \\
  13.45 &   14.26 &   12.53 &   11.3 &   10.4 &   10.0 &                V499-Ori \\
  13.47 &   17.06 &   11.39 &    7.3 &    6.3 &    5.8 & 2MASS-J12450626-7720133 \\
  13.50 &   14.76 &   12.35 &   10.6 &    9.8 &    9.3 &                   SZ-72 \\
  13.52 &   14.36 &   12.59 &   11.1 &   10.3 &    9.8 &               Haro-5-34 \\
  13.54 &   15.13 &   12.13 &   10.0 &    8.9 &    8.4 &         IRAS-12535-7623 \\
  13.66 &   14.43 &   12.79 &   11.5 &   10.6 &   10.0 & 2MASS-J05394017-0220480 \\
  13.66 &   14.49 &   12.75 &   11.7 &   10.8 &   10.2 &                V543-ORI \\
  13.95 &   15.45 &   12.67 &   10.2 &    8.7 &    7.6 & 2MASS-J16394544-2402039 \\
\hline
%\end{tabular}
\end{longtable}
\twocolumn
\end{appendix}

\end{document}